\documentclass[apj]{emulateapj}
\usepackage{natbib}
\shorttitle{The VMC Survey. XVIII.}
\shortauthors{C. Zhang et al.}
\usepackage{color}

\begin{document}

\title{The VMC Survey. XVIII. Radial dependence of the low-mass,
  0.55--0.82 $M_\odot$ stellar mass function in the Galactic globular
  cluster 47 Tucanae}

\author{
Chaoli Zhang\altaffilmark{1,2}
Chengyuan Li\altaffilmark{1,3,4},
Richard de Grijs\altaffilmark{1,3,5},
Kenji Bekki\altaffilmark{6},
Licai Deng\altaffilmark{7},
Simone Zaggia\altaffilmark{8},
Stefano Rubele\altaffilmark{8},
Andr\'es E. Piatti\altaffilmark{9,10},
Maria-Rosa L. Cioni\altaffilmark{11,12,13},
Jim Emerson\altaffilmark{14},
Bi-Qing For\altaffilmark{7},
Vincenzo Ripepi\altaffilmark{15},
Marcella Marconi\altaffilmark{15}, 
Valentin D. Ivanov\altaffilmark{16},
and 
Li Chen\altaffilmark{2} 
}

\altaffiltext{1} {Kavli Institute for Astronomy \& Astrophysics,
  Peking University, Yi He Yuan Lu 5, Hai Dian District, Beijing
  100871, China; jackzcl@outlook.com, grijs@pku.edu.cn}
\altaffiltext{2} {Shanghai Astronomical Observatory, Chinese
Academy of Sciences, Shanghai 200030, China}  
\altaffiltext{3} {Department of Astronomy, Peking University, Yi He
  Yuan Lu 5, Hai Dian District, Beijing 100871, China}
\altaffiltext{4} {Purple Mountain Observatory, Chinese Academy 
of Sciences, BeiJing Xi Lu, Nanjing 210008, China}  
\altaffiltext{5} {International Space Science Institute--Beijing, 1
  Nanertiao, Zhongguancun, Hai Dian District, Beijing 100190, China}
\altaffiltext{6} {ICRAR M468, The University of Western Australia, 35
  Stirling Highway, Crawley, WA, 6009, Australia}   
\altaffiltext{7} {Key Laboratory for Optical Astronomy, National
  Astronomical Observatories, Chinese Academy of Sciences, 20A Datun
  Road, Chaoyang District, Beijing 100012, China} 
\altaffiltext{8} {INAF-Osservatorio Astronomico di Padova, vicolo
  dell'Osservatorio 5, I-35122 Padova, Italy} 
\altaffiltext{9} {Observatorio Astro\'nomico, Universidad Nacional de
  C\'ordoba, Laprida 854, 5000, C\'ordoba, Argentina}
\altaffiltext{10} {Consejo Nacional de Investigaciones
  Cient\'{\i}ficas y T\'ecnicas, Av. Rivadavia 1917, C1033AAJ, Buenos
  Aires, Argentina}
\altaffiltext{11} {Universit\"{a}t Potsdam, Institut f\"{u}r Physik 
und Astronomie, Karl-Liebknecht-Str. 24/25, 14476 Potsdam, Germany}   
\altaffiltext{12} {Leibnitz-Institut f\"ur Astrophysik Potsdam, An der
  Sternwarte 16, D-14482 Potsdam, Germany} 
\altaffiltext{13} {Physics, Astronomy, and Mathematics,
  University of Hertfordshire, Hatfield AL10 9AB, UK}   
\altaffiltext{14} {Astronomy Unit, School of Physics and Astronomy,
  Queen Mary University of London, Mile End Road, London E1 4NS, UK}
\altaffiltext{15} {INAF-Osservatorio Astronomico di Capodimonte, via
  Moiariello 16, 80131 Naples, Italy}
\altaffiltext{16} {European Southern Observatory,
  Karl-Schwarzschild-Str. 2, Garching bei M\"unchen, 85748, Germany}

\begin{abstract}
We use near-infrared observations obtained as part of the {\sl Visible
  and Infrared Survey Telescope for Astronomy} (VISTA) Survey of the
Magellanic Clouds (VMC), as well as two complementary {\sl Hubble
  Space Telescope} ({\sl HST}) data sets, to study the luminosity and
mass functions as a function of clustercentric radius of the
main-sequence stars in the Galactic globular cluster 47 Tucanae. The
{\sl HST} observations indicate a relative deficit in the numbers of
faint stars in the central region of the cluster compared with its
periphery, for $18.75\leq m_{\rm F606W}\leq 20.9$ mag (corresponding
to a stellar mass range of $0.55<m_\ast/{M_\odot}<0.73$). The stellar
number counts at $6.7'$ from the cluster core show a deficit for
$17.62\leq m_{\rm F606W}\leq 19.7$ mag (i.e.,
$0.65<m_\ast/{M_\odot}<0.82$), which is consistent with expectations
from mass segregation. The VMC-based stellar mass functions exhibit
power-law shapes for masses in the range $0.55<m_\ast/{M_\odot}<
0.82$. These power laws are characterized by an almost constant slope,
$\alpha$. The radial distribution of the power-law
slopes $\alpha$ thus shows evidence of the importance of both mass
segregation and tidal stripping, for both the first- and
second-generation stars in 47 Tuc.
\end{abstract}

\section{Introduction}
\label{sec:intro}

The globular clusters (GCs) in the Milky Way are excellent probes to
study the Galaxy's formation history. One of the major goals in
contemporary astrophysics is to understand the stellar mass functions
(MFs) of GCs, because they are thought to contain original information
about the stellar initial mass function (IMF). A number of physical
processes can cause the IMF to vary, including fragmentation,
accretion, feedback, stellar interactions, and magnetic-field
contributions \citep{Larson:1992aa,Padoan:2002aa,Bonnell:2007aa}. A
detailed understanding of the IMF is required to study the stellar
populations of external galaxies, which would otherwise be impossible
to study owing to their unresolved nature.

However, it is not simple to derive the IMF from an observed
present-day MF, since the latter is affected by both observational and
theoretical limitations: observationally, it is significantly impaired
by the crowding of GCs, photometric uncertainties, and the
observational field of view, which can lead to significant biases when
attempting to determine the underlying stellar luminosity function
\citep[LF;][]{King:1958aa}. The situation is further complicated by
the fact that the mass--luminosity relation (MLR), which is required
to transform the observed stellar luminosities to the corresponding
masses, is metallicity- and age-dependent
\citep[e.g.,][]{Kroupa:1990aa,de-Grijs:2002aa}. Theoretical MLRs for
intermediate and high stellar masses ($m_\ast \ge 0.2 M_{\odot}$) are
relatively well constrained observationally, but at the low-mass end
($m_\ast \le 0.2 M_{\odot}$) significant uncertainties remain. In
addition, the MF can also be modified by dynamical processes over a
cluster's lifetime. For instance, tidal interactions with the Milky
Way's gravitational potential affect the outskirts of clusters
\citep{Lane:2012aa}, while mass segregation affects their central
regions \citep[e.g.,][]{de-Grijs:2002aa}.

Despite all these observational and theoretical challenges, the MF of
the Galactic GC 47 Tucanae (47 Tuc) has been studied
extensively. \cite{Paust:2010aa} showed that the 47 Tuc MF in the
cluster's central region can be approximated by a 
power-law distribution, i.e., ${\rm d}N/{\rm d}M\propto M^{-\alpha}$,
with $\alpha=0.84$ for the stellar mass range of 0.2--$0.8 M_{\odot}$;
\cite{de-Marchi:1995aa} and \cite{Santiago:1996aa} showed that the MF
at a location between $4'$ and $5'$ from the center (close to the
cluster's half-mass radius) is a power law with
$\alpha=1.5$ for the mass range of 0.3--0.8 $M_{\odot}$, but that it
flattens in the range 0.14--0.3 $M_{\odot}$. Comparison with the most
recent Monte Carlo simulations by \cite{Giersz:2011aa} showed that the
stars above the main-sequence (MS) turn-off in 47 Tuc obey a power-law
with $\alpha=2.8$ and follow a relatively flat IMF with an index of
about 0.4 along the lower MS. This mass distribution is much shallower
than that found based on the ground-based observations of
\cite{Hesser:1987aa}, who determined an index of 1.2 for the mass
range of 0.5--0.8 $M_{\odot}$. However, all stellar MFs investigated
in previous studies pertained to specific loci in the cluster; some
were located in the cluster center, some were based on data from its
periphery, and some were determined around the half-mass radius. Here
we present a systematic analysis covering a much larger radial range
using ground-based data obtained with the 4m {\sl Visible and Infrared
  Survey Telescope for Astronomy} (VISTA) telescope as part of the
VISTA Survey of the Magellanic Clouds (VMC), combined with {\sl HST}
observations at central and intermediate cluster radii, to investigate
the low-mass MS LF and MF as a function of radius from the center of
47 Tuc.

This paper is organized as follows. In Section 2, we present the
observational data and our analysis method. Our results are presented
in Section 3. In Section 4, we discuss our results in the context of
mass segregation and tidal stripping processes affecting 47 Tuc. Our
conclusions are provided in Section 5.

\section{Data Selection and Analysis}
\label{sec:data}

\subsection{VISTA data}
\label{ph: VMC}

\begin{figure}[]
\begin{center}
\includegraphics[totalheight=0.4\textheight]{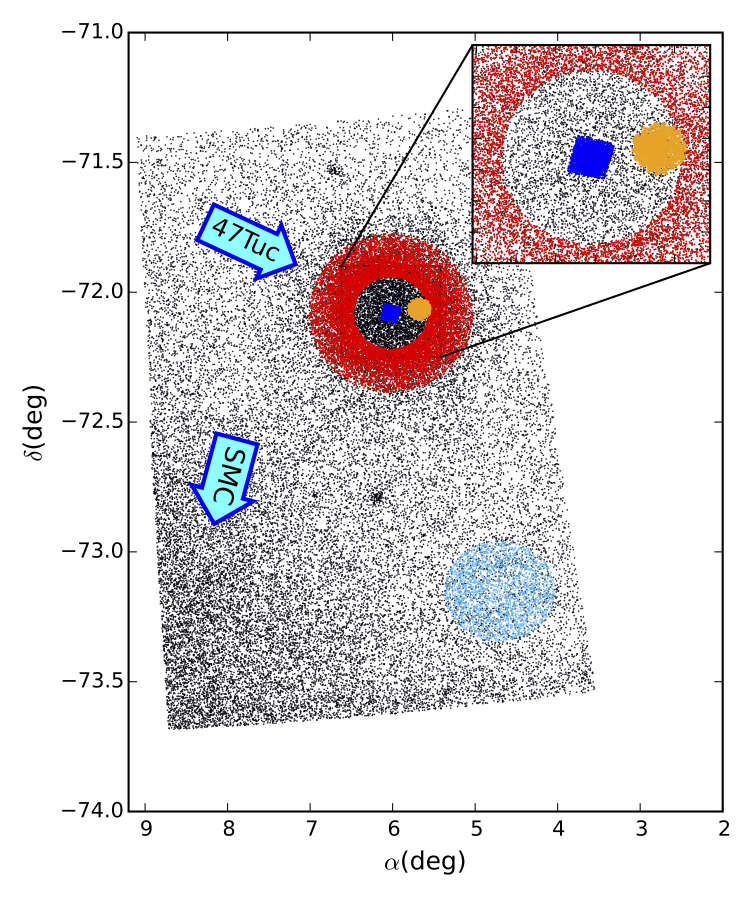}
\end{center}
\caption{Spatial distribution of the stars in 47 Tuc (coordinates are
  given for the J2000 epoch), combining SMC tiles 5$\_$2 (top half)
  and 4$\_$2 (bottom half). The background stars were drawn from the
  VISTA data; the red annulus indicates the VMC data used for this
  study, which is further radially binned into five subsets. The blue
  region corresponds to the {\sl HST} catalog of Sarajedini et
  al. (2007), and the orange region represents the ultra-deep {\sl
    HST} catalog of Kalirai et al. (2012). The cyan region is adopted
  to compute the background field-star density, which is in turn used
  to decontaminate the cluster CMD.}
\label{Fig:Spatial_distri}
\end{figure}	

The near-infrared $Y$, $J$, and $K_\mathrm{s}$ observations used in
this study were obtained as part of the VMC survey
\citep{Cioni:2011aa}. 47 Tuc is projected onto the Small Magellanic
Cloud (SMC) and is located on VMC tile SMC 5$\_$2. The SMC's main body
is partially located on the adjacent, southern tile SMC 4$\_$2. The
VMC data were processed with the VISTA Data Flow System pipeline
(VDFS) and calibrated to match the VISTA photometric system, which is
close to the Vegamag system \citep{Irwin:2004aa}. We extracted the
paw-print VMC images for all three filters from the VISTA Science
Archive and used the {\sc iraf}/{\sc daophot} package to derive the
point-spread functions \citep[PSFs;][]{Stetson:1987aa}. The data
selected for this study, composed of two epochs in the $Y$ and $J$
bands and nine in $K_\mathrm{s}$, were PSF-homogenized and stacked to
obtain a final, deep tile image.\footnote{We found that the VMC
  observations of 47 Tuc are significantly affected by PSF and
  sensitivity inhomogeneities for some epochs. For this reason, we
  selected only epochs characterized by minimal seeing variations and
  the highest sensitivity for analysis in the present paper, resulting
  in two epochs in the $Y$ band, two and one concatenation in $J$
  (corresponding to two and a half epochs), and nine in $K_{\rm s}$.}
We performed PSF photometry on the homogenized, deep VMC SMC 5$\_$2
and 4$\_$2 tiles image using the {\sc psf} and {\sc allstar}
tasks. The photometry was calibrated using the method described in
\cite{Rubele:2015aa}. Subsequently, we correlated the three-band
photometry to produce a multi-band catalog. Figure
\ref{Fig:Spatial_distri} shows the spatial distribution of the MS
stars in 47 Tuc. The observations cover an area of almost $2\times5$
deg$^{2}$ (one VISTA InfraRed Camera -- VIRCam -- tile consists of six
offset paw-print exposures covering a $\sim 1.6$ deg$^{2}$ field of
view), which contains more than 190,000 stars with $Y \in [16.8,19.6]$
mag. A detailed overview of our observations of 47 Tuc as well as of
our analysis procedures, was given by \cite{Li:2014aa}.

The cluster is located toward the southwest in the SMC 5$\_$2 tile
image, while the southeastern corner of the field is dominated by SMC
field stars. Therefore, the area occupied by 47 Tuc could contain
significant numbers of field stars associated with both the SMC and
the Milky Way. SMC tile 5$\_$2 cannot be used for the field-star
decontamination, since the cluster's tidal radius can reach $r_{\rm t}
= 2500''$ \citep{Harris:1996aa,Lane:2012aa}, which essentially covers
the entire tile. For background star decontamination, it is crucial to
find a region that is representative both in terms of the star counts
from the Milky Way and their counterparts from the variable SMC
background. The bottom right-hand corner of the field, shown in cyan
in Fig. \ref{Fig:Spatial_distri} and contained in SMC tile 4$\_$2, is
ideal for our decontamination of the cluster's color--magnitude
diagram (CMD) from background stars, for two reasons. First, it is
located at the same Galactic latitude as 47 Tuc so that Galactic field
contamination should be similar to that affecting our 47 Tuc
observations; second, this region is located at a suitable distance
from both 47 Tuc and the SMC, thus minimizing the number of possible
residual 47 Tuc and SMC stars.

We therefore adopted a region dominated by the 47 Tuc member stars
centered at $\alpha_{\rm J2000} =00^{\rm h}24^{\rm m}04.80^{\rm
  s}(6.020^{\circ})$, $\delta_{\rm J2000} = -72^{\circ}04'48''
(-72.080^{\circ})$ and within a radius of $1100''$
\citep{Li:2014aa}. A second region, shown in cyan in
Fig. \ref{Fig:Spatial_distri}, centered on $\alpha_{\rm J2000}
=00^{\rm h}18^{\rm m}48.00^{\rm s} (4.70^{\circ})$, $\delta_{\rm
  J2000} = -73^{\circ}09'02''(-73.15^{\circ})$ with a radius of
$600''$, was adopted to calculate the background stellar density.

\cite{Li:2014aa} showed, for the same VMC data, that the observational
completeness levels drop off rapidly in the cluster's inner region.
This is caused by a combination of the increased blending probability
and the enhanced background brightness. They performed a large number
of artificial-star tests to study the effects of crowding on the
uncertainties in the resulting PSF photometry. For details of the
artificial-star tests, please refer to \cite{Rubele:2012aa}. The
latter authors generated $\sim10^{6}$ artificial stars in the image
and repeated their PSF photometry in the same manner as for our sample
of real stars. This resulted in an artificial-star catalog that
contained the input and output magnitudes, as well as the photometric
errors, computed as ``output minus input'' magnitudes. Our
observations' completeness reaches a level of 50\% at $Y = 19.6$ mag
within a radius of $500''$ from the cluster center. We therefore
restricted our study to MS stars in an annulus defined by $r \in
[500,1100]''$ and $Y \in [16.8,19.6]$ mag.

\subsection{Hubble Space Telescope data}
\label{ph: HST}

We also made use of two different {\sl HST} data sets of 47 Tuc to
study the cluster's central regions, $r \in [0,500]''$, which are not
resolved by the VMC data. Both data sets were obtained with {\sl
  HST}'s Advanced Camera for Surveys (ACS). One of the data sets was
taken as part of the Globular Cluster Treasury program \citep[PI:
  A. Sarajedini;][hereafter SA07]{Sarajedini:2007aa}, which aimed at
obtaining accurate photometry for stars well below the MS turn-off in
the F606W and F814W filters. We directly use the
\cite{Anderson:2008aa} catalog: see the blue data points in
Fig. \ref{Fig:Spatial_distri}.

Second, we used observations obtained by \cite[][hereafter
  KA12]{Kalirai:2012aa}, who collected photometry for white dwarfs in
47 Tuc based on 121 orbits of Cycle 17 {\sl HST} observations. Their
main goal was to obtain photometry in the ACS F606W and F814W filters
with extremely long exposure times to reach very faint magnitudes
(approaching 29 mag in F606W) to study the entirety of the white dwarf
cooling sequence in the cluster. The advantages of using the KA12 data
are, first, that their field is located $6.7'$ (8.8 pc) west of the
cluster center. This region is neither too sparse nor too crowded for
our analysis. It allows us to resolve large numbers of MS stars down
to very low luminosities. Second, the authors' completeness tests
demonstrate that their photometry is very precise and place the 50\%
completeness limit for the F606W filter at 29.75 mag. A detailed
discussion of the observations and the corresponding data reduction
can be found in KA12. The KA12 catalog data points are colored orange
in Fig. \ref{Fig:Spatial_distri}.

In the context of the study presented here, we emphasize that for both
{\sl HST} data sets, we restricted our study to magnitudes of $m_{\rm
  F606W} \in [17.53,20.9]$ mag for comparison with the VMC data. The
{\sl HST} data was decontaminated by proper-motion selection. This
thus enabled us to construct highly robust local LFs.

\subsection{Isochrone fitting and MS selection}
\label{ph: cmdfit}

\begin{table}
 \centering
 \begin{minipage}{85mm}
  \caption{Basic representative 47 Tuc CMD fit parameters.}
  \label{tab: cmdpara}
  \begin{tabular}{@{}lccc@{}}
  \hline
Parameter						& {\sl HST} SA07 		& {\sl HST} KA12 	& VISTA VMC\tabularnewline
\hline
Model 						& DSEP\footnote{DSEP: \cite{Dotter:2008aa}} 	& DSEP         
						 	& PGPUC\footnote{PGPUC: \cite{Valcarce:2012aa} } \tabularnewline
$t$ (Gyr) 						& 12.5 				& 12.5 		 	& 12.5\tabularnewline
$Y$ 							& 0.251 				& 0.252 			& 0.26 \tabularnewline
$[{\rm Fe/H}]$ (dex) 			 		& $-$0.66 				& $-$0.74 			& $-$0.83\tabularnewline
$[\alpha/{\rm Fe}]$ (dex) 			& 0.0 				& 0.4 			& 0.3 \tabularnewline
$E(B-V)$ (mag)\footnote{\cite{Harris:1996aa}} & 0.04 & 0.04 				& 0.04 \tabularnewline
$(m-M)_0$ (mag) 				& 13.3 				& 13.3 			& 13.35\tabularnewline
$\alpha_{{\rm J2000}}$ & $00^{{\rm h}}24^{{\rm m}}04.80^{{\rm s}}$ & $00^{{\rm h}}22^{{\rm m}}39.00^{{\rm s}}$& -- \tabularnewline
$\delta_{{\rm J2000}}$  & $-72^{\circ}04'48''$ & $-72^{\circ}04'04''$ & -- \tabularnewline
MS range (mag)  				& $[17.53, 20.9]$ 		&  $[17.62, 20.9]$ 	& $ [16.8,19.6]$ \tabularnewline
Completeness (\%) 				& $[99, 85]$ 			&  $[100, 99.8]$ 	& $[86, 65]$\footnote{For stars in an annulus with radii of 500$''$ and 1100$''$} \tabularnewline

\hline 
\end{tabular}
\end{minipage}
\end{table}	

\begin{figure*}[]
\begin{center}
\includegraphics[totalheight=0.4\textheight]{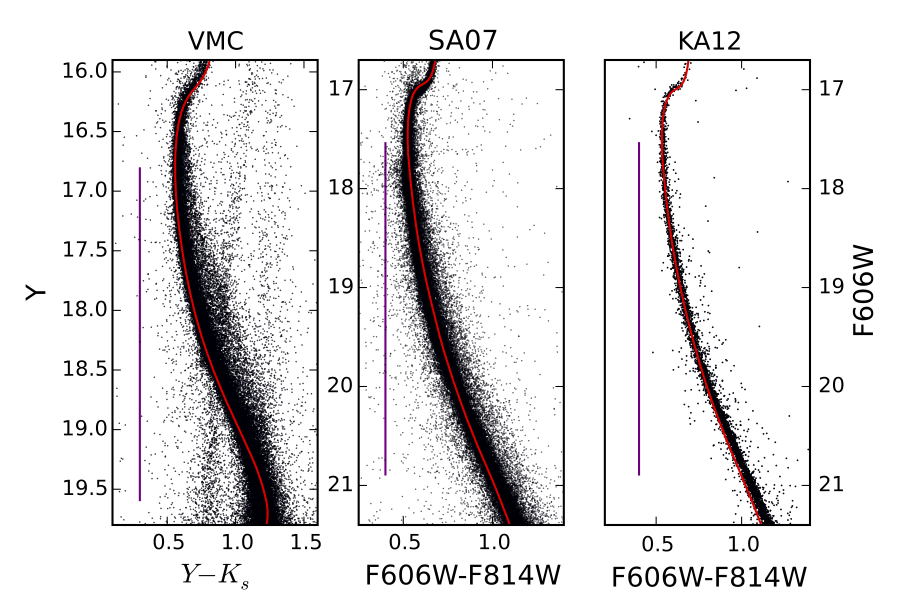}
\end{center}
\caption{47 Tuc CMDs. The red curves are the model isochrones defined
  in Table \ref{tab: cmdpara}. The vertical purple lines indicate the
  magnitude ranges adopted for selecting MS stars.}
\label{Fig:CMD}
\end{figure*}

A number of studies have demonstrated that 47 Tuc displays multiple
stellar populations across its entire CMD
\citep[e.g.,][]{Anderson:2009aa,Milone:2012ab,Li:2014aa}. We therefore
adopted slightly different isochrones to fit the CMDs corresponding to
the different data sets used in this paper. Table \ref{tab: cmdpara}
summarizes their basic parameters. The metallicity ([Fe/H]) adopted in
the table decreases slightly (from [Fe/H] = $-$0.66 dex to [Fe/H] =
$-$0.82 dex) with increasing radius, which is consistent with the
results of \cite{Li:2014aa}, who found that stars in the cluster core
are more metal-rich than their counterparts in the cluster's
periphery. The theoretical isochrones we used to fit our CMDs were
taken from the Dartmouth Stellar Evolution Program
\cite[DSEP;]{Dotter:2008aa} for the {\sl HST} data sets, since the
DSEP models are highly commensurate with other models
\cite[e.g.,][]{Pietrinferni:2006aa} and thus provide a robust and
reliable MLR. However, the DSEP suite does not provide models in the
VISTA $Y$, $J$, and $K_\mathrm{s}$ photometric system. Therefore, we
adopted the Princeton--Goddard--Pontificia Universidad Cat\'olica
(PGPUC\footnote{http://www2.astro.puc.cl/pgpuc/iso.php}) stellar
evolutionary code \citep{Valcarce:2012aa} to obtain best fits to the
VMC data. The red curves in Figure \ref{Fig:CMD} shows the CMDs and
their corresponding best-fitting isochrones; the vertical purple lines
in all panels show the magnitude ranges adopted for selecting the
corresponding MS regime.


\begin{figure}[]
\begin{center}
\includegraphics[totalheight=0.25\textheight]{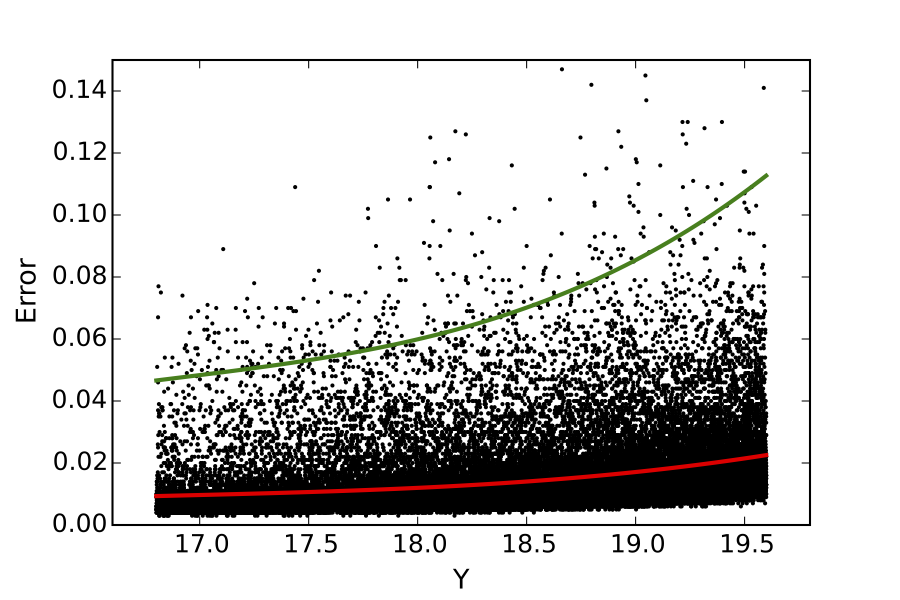}
\end{center}
\caption{Photometric uncertainties as a function of magnitude for the
  VMC data set in the $Y$ filter. The red solid curve represents the
  bin-averaged photometric uncertainties, while the green curve
  represents the $5 \sigma$ range.}
\label{Fig: error}
\end{figure}

\begin{figure*}[]
\begin{center}
\includegraphics[totalheight=0.4\textheight]{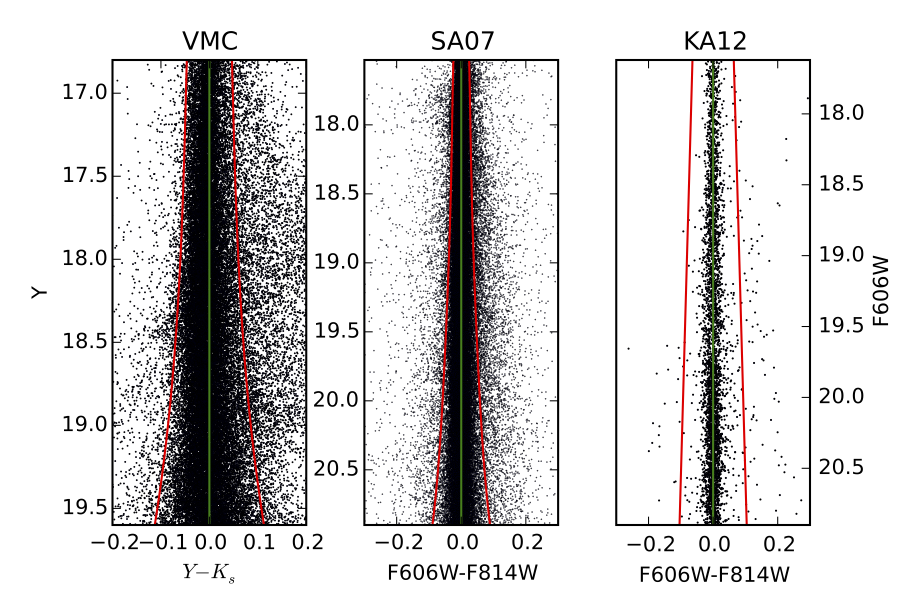}
\end{center}
\caption{47 Tuc CMD, ``normalized'' by removing the trend defined by
  the MS ridge line. The red lines are the $5 \sigma$, $4 \sigma$, and
  $5 \sigma$ selections used to correct for the presence of background
  field stars. The green line is the zero-color reference. }
\label{Fig: fiducial_plot}
\end{figure*}

In order to select all MS stars in 47 Tuc with minimal contamination
owing to photometric uncertainties, we characterized the photometric
errors in each passband and each data set as follows. We first divided
the magnitude range in each band into 100 bins and subsequently
determined the mean photometric uncertainty in each bin. Subsequently,
we interpolated the bin-averaged values. Figure \ref{Fig: error} shows
an example for the VISTA $Y$ filter: the bin-averaged photometric
uncertainties are plotted as the red curve, while the green curve
represents the $5\sigma$ range. We also determined the fiducial ridge
line of the MS stars in the CMDs, using bin sizes of 0.3 mag for $Y
\in [16.8,19.6]$ mag and $m_{F606W} \in [17.53,20.9]$ mag. The
fiducial ridge lines for all three data sets were then used to
normalize the CMDs: see Fig. \ref{Fig: fiducial_plot}. In this figure,
the red curves indicate all stars selected within pre-determined
photometric uncertainty ranges, i.e., $5 \sigma$, $4 \sigma$, and $5
\sigma$ for SA07, KA12, and VMC data, respectively. This selection is
required to limit unmodeled effects owing to the presence of a
population of unresolved binary systems.

\section{Results: }
\label{sec: results}

\subsection{Luminosity functions}

\begin{figure*}[]
\begin{center}
\includegraphics[totalheight=0.5\textheight]{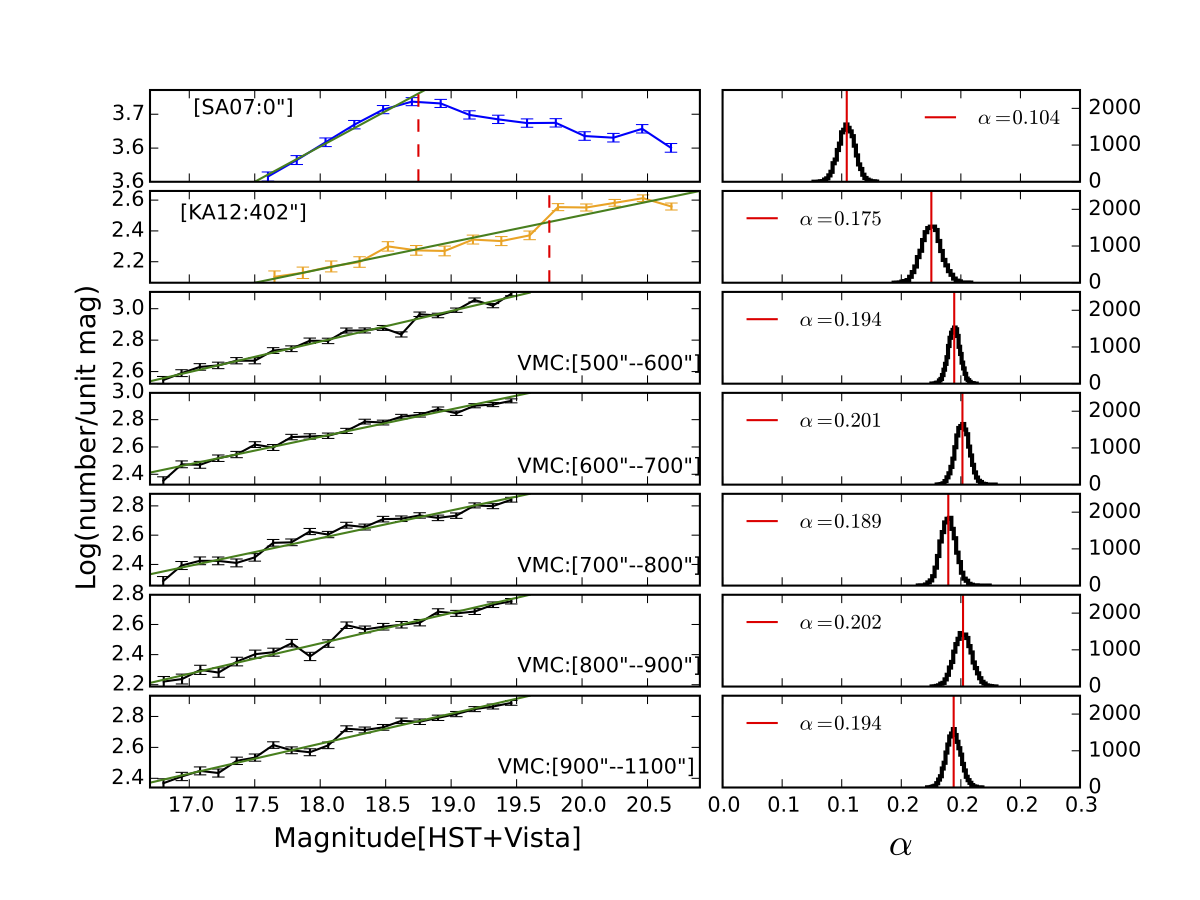}
\end{center}
\caption{47 Tuc MS LFs. LFs based on (top) the SA07 catalog for stars
  in the cluster center and (second panel) the KA12 catalog,
  containing stars located $6.7'$ west of the cluster center. The
  bottom five panels are local LFs for radial annuli covered by our
  VMC observations. The full annulus, $r \in [500,1100]''$, is divided
  into five radial subsets, i.e., $r \in [500,600]''$, $r \in
  [600,700]''$, $r \in [700,800]''$, $r \in [800,900]''$, and $r \in
  [900,1100]''$. The numbers of MS stars per unit magnitude versus
  apparent magnitude F606W are plotted; the error bars were estimated
  based on Poissonian counting statistics. The green line is the
  power-law fit to the LF, and the two arrows in top two panels are an
  indication of a possible LF break magnitude. The Gaussian
  distributions on the right are the results of our Monte Carlo
  simulations to determine the best-fitting power-law index $\alpha$.}
\label{Fig: lum}
\end{figure*}

The completeness-corrected local LFs based on the two {\sl HST} data
sets are shown in the top panel of Fig. \ref{Fig: lum}. The stars in
both CMDs were divided into 15 bins. The error bars represent
Poissonian counting statistics. The LFs extend over the magnitude
range $17.53\leq m_{\rm F606W} \leq 20.9$ mag. However, we note that
the SA07 LF declines toward the low-mass end, for $18.75\leq m_{\rm
  F606W}\leq 20.9$ mag, while the KA12 LF exhibits a deficit at the
high-mass end, $17.62\leq m_{\rm F606W}\leq19.7$ mag. The vertical
dashed lines in both panels indicate the decline on the left of the
dashed line for SA07 and the deficit on the right for the KA12 data.

We consequently truncated the LF based on the SA07 data at the
luminosity where the linear regime (indicated in green) is fitted
adequately by a power law, i.e. ${\rm d}N(L)/{\rm d}L
\propto L^{-\alpha}$, where $N(L)$ corresponds to the number of stars
per unit luminosity $L$. To determine $\alpha$ statistically robustly,
we ran Monte Carlo simulations, assuming that the stellar number
counts in each bin are well-represented by Gaussian distributions. We
then randomly drew stellar number counts from each bin to construct
our sample LFs and fitted the results using a power law to obtain
$\alpha$. This random sampling based on Gaussian distributions in each
bin was repeated $\sim 15,000$ times. The resulting $\alpha$
distributions are shown in the right-hand panels of \ref{Fig: lum},
leading to (mean) $\mu_{\alpha}=0.104$ for the SA07 data and
$\mu_{\alpha}=0.175$ for the KA12 catalog.

The VMC-based LF of 47 Tuc is shown at the bottom of Fig. \ref{Fig:
  lum}, also corrected for the effects of sampling incompleteness and
field-star contamination. We divided the annulus $r \in [500,1100]''$
further into five radial subsets, i.e., $r \in [500,600]$, $r \in
[600,700]''$, $r \in [700,800]''$, $r \in [800,900]''$, and $r \in
[900,1100]''$. Applying the same routines as for the {\sl HST} data,
for each annulus we fitted the observed LFs with a power
law and ran Monte Carlo simulations to obtain the distribution of
$\alpha$. The results showed that the LFs at different radii are all
closely approximated by power laws, and that the $\alpha$ index
increases toward the outer regions of the cluster.

\subsection{Mass--luminosity relation and mass functions}

\begin{figure}[]
\begin{center}
\includegraphics[totalheight=0.25\textheight]{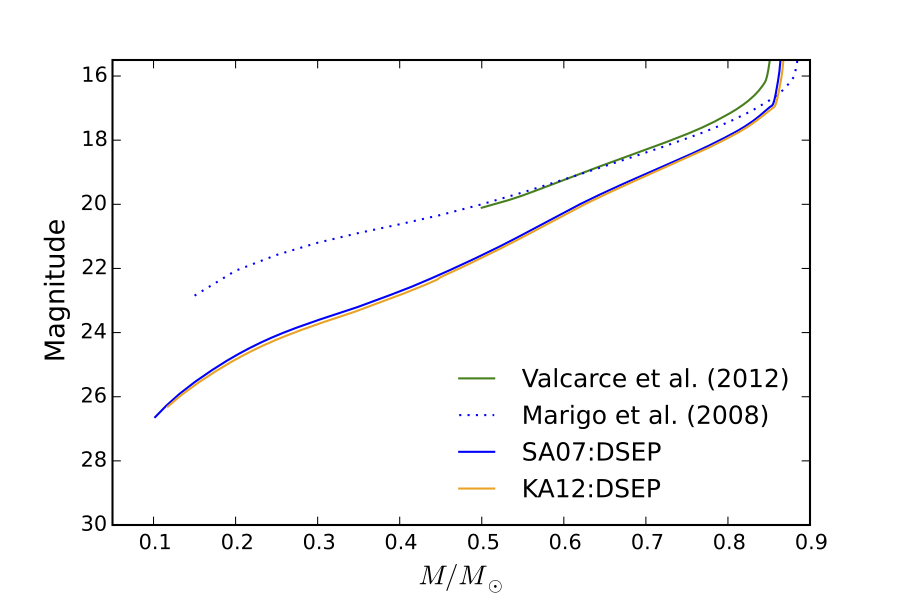}
\end{center}
\caption{MLR based on the DSEP, PGPUC \citep{Valcarce:2012aa}, and
  Padova \citep{Marigo:2008aa,Girardi:2010aa} models for both the
  VMC's $Y$-band magnitudes and the {\sl HST} F606W filter. The blue
  and orange curves are the DSEP MLRs for the {\sl HST} data, while
  the green curve represents the PGPUC model for the VMC data. The
  dotted line is the Padova model pertinaing to the VMC data, shown
  for comparison. }
\label{Fig: MLR}
\end{figure}

To determine the stellar mass function (MF), the observed LF is
commonly divided by the derivative of the MLR. However, the shape of
the MLR depends sensitively on the stellar model used. Figure
\ref{Fig: MLR} includes the three most-up-to-date stellar evolution
models---DSEP \citep{Dotter:2008aa}, PGPUC \citep{Valcarce:2012aa},
and the Padova models \citep{Marigo:2008aa,Girardi:2010aa}---in both
the VISTA photometric system and for the {\sl HST}/ACS photometric
passbands used by SA07 and KA12. The stellar mass range we study here
is $0.55 \leq m_\ast/{M_\odot} \leq 0.82$, a range where the input
physics is relatively better constrained than at the low-mass end, and
therefore all models are characterized by smooth MLRs. We adopted
PGPUC and DSEP as our reference models.
	
\begin{figure*}[]
\begin{center}
\includegraphics[totalheight=0.5\textheight]{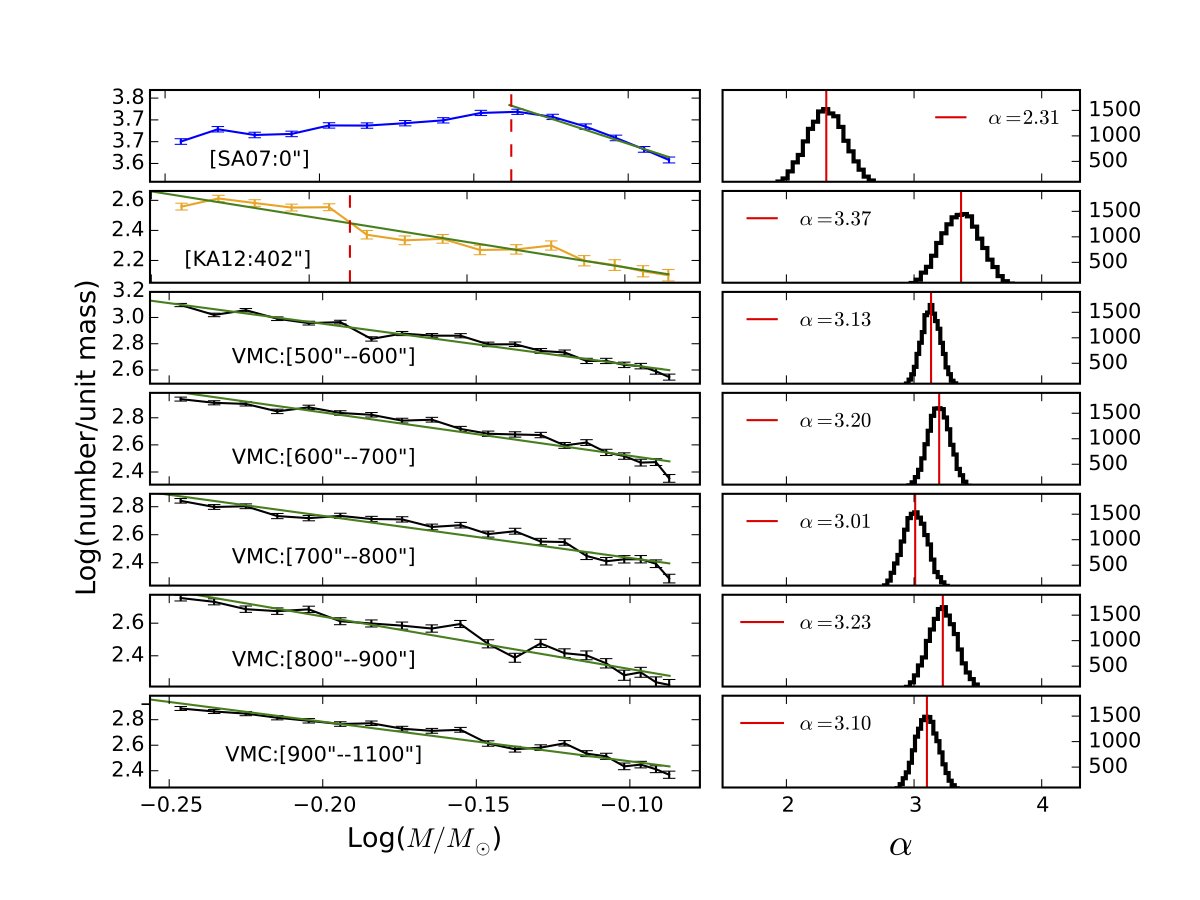}
\end{center}
\caption{47 Tuc MFs, obtained by dividing the slope of the MLR by
  the LFs of Fig. \ref{Fig: lum}. Top two panels: The SA07 MF,
  pertaining to the cluster's central region, shows a decline in
  stellar numbers for $m_\ast < 0.72 M_{\odot}$. In contrast, the
  KA12-based MF in the outer region reveals a deficit for $m_\ast >
  0.65 M_{\odot}$. Bottom five panels: MFs at different radii in the
  outskirts of 47 Tuc. All annuli exhibit power laws, with no evidence
  of any deficit or surplus for masses in excess of 0.6
  $M_{\odot}$. We thus fitted all the MFs with the power
  laws and performed Monte Carlo simulations, shown on the right.}
\label{Fig: mass}
\end{figure*}

The left-hand panels of Fig. \ref{Fig: mass} show the 47 Tuc MF for
our two {\sl HST} data sets and for the VMC data. The Monte Carlo
simulation results for the $\alpha$ distribution of power-law fits are 
indicated in the right-hand panels. As already
shown for the LFs, the SA07 MF shows a decline toward the low-mass
end, for a mass range of $0.55<m_\ast/{M_\odot}<0.73$, whereas a
deficit is again observed in the KA12 data for the high-mass end, at
$0.65<m_\ast/{M_\odot}< 0.82$ (see the red dashed lines). Similarly to
our approach pertaining to the LFs, we truncated the SA07 MF where it
starts to deviate from a power-law distribution and fitted a power law
to this part of the MF. The resulting $\alpha$ distribution increases
from $\alpha=2.31$ (SA07; center) to $\alpha=3.37$ (KA12; $6.7'$ from
the center), this result is in line with the expectations from mass
segregation.

The local MFs based on our VMC data are shown at the bottom of
Fig. \ref{Fig: mass} for different clustercentric radii. The MFs are
all power laws and exhibit almost constant $\alpha$ indices.  The mean
value of the five indices is $\langle \alpha \rangle = 3.13$, with a
standard derivation of $\sigma_{\alpha} = 0.07$. We will discuss this
flat trend in $\alpha$ index in the next section in the context of the
effects of mass segregation and tidal stripping. For
  comparison, \cite{Bochanski:2010aa} found $\alpha = 2.38$ for the MF
  of low-mass dwarfs in the field over the mass range $0.32
  M_{\odot}<m_\ast<0.8 M_{\odot}$, while \cite{Kalirai:2013aa}
  determined that the field MF in the outer regions of the SMC for the
  stellar mass range $m_\ast \in [0.37,0.93] M_{\odot}$ is well
  presented by a power-law with $\alpha=1.9$. Except for the innermost
  annulus, our 47 Tuc MFs are thus steeper than expected for field
  stars, which reflects the effects of dynamical processing.
	
\section{Discussion}
\label{sec: discussion}

\begin{figure}[]
\begin{center}
\includegraphics[totalheight=0.25\textheight]{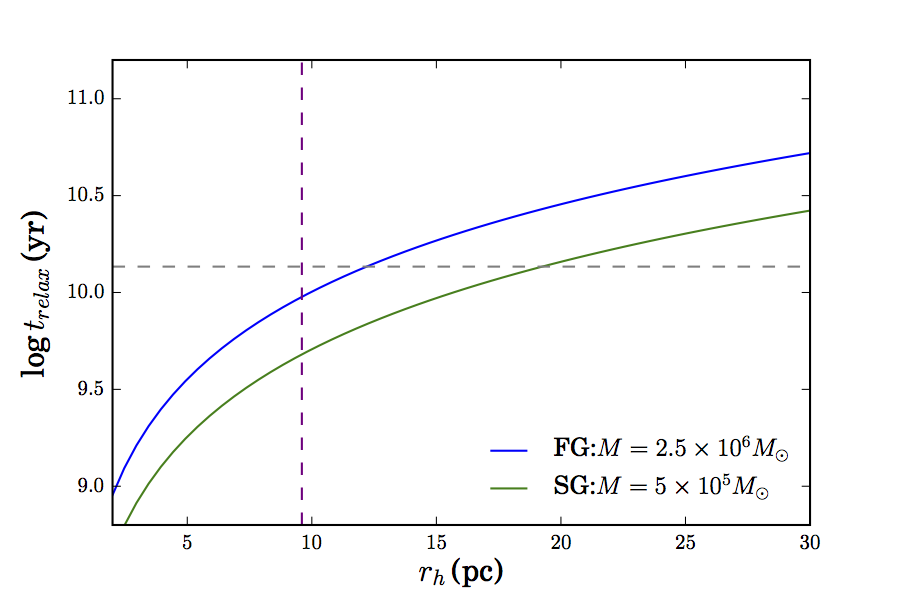}
\end{center}
\caption{Half-mass relaxation timescale for (blue) first-generation
  stars and (green) second-generation stars as a function of radius in
  47 Tuc. The grey dashed line represents an age of 13 Gyr, which
  intersects the first-generation prediction at $r = 12.3$ pc and the
  second-generation curve at $r = 19.1$ pc. The purple dashed line
  represents the observed half-mass radius at $r = 9.6$ pc.}
\label{Fig: timescale}
\end{figure}	

The observed MF $\alpha$ index increases from 2.31 in the central
region (SA07) to 3.37 at intermediate radii (at $r \approx 6.7'$;
KA12). This shows that mass segregation significantly affects the
resulting stellar MF as a cluster undergoes dynamical relaxation
\citep{Anderson:1996aa}. Dynamical mass segregation causes the
high-mass stars to attain lower velocities, and they will thus
gradually sink toward the cluster center and boost the lower-mass
stars to the cluster's periphery. This will, in turn, lead to a
flatter MF in the inner regions and a gradually steeper MF when moving
toward the outskirts. This is evident for 47 Tuc, given the observed
declining MF trend for low-mass stars in the central region, and the
clear drop in stellar numbers starting from a mass of $m_\ast = 0.65
{M_\odot}$ or $\log(m_\ast/{M_\odot}) = -0.18$ dex. This drop cannot
have been caused by incompleteness effects nor by field-star
contamination, since the KA12 catalog is proper-motion cleaned and is
100\% complete at the brightness levels of interest here. The radial
diffusion of low-mass stars was recently observed for the first time
by \cite{Heyl:2015aa}, based on a sample of young white dwarfs.

This dynamical evolution can be further complicated by the recent
discovery that 47 Tuc hosts multiple stellar populations
\citep{Milone:2012ab,Li:2014aa}. \cite{DErcole:2010aa} and
\cite{Bekki:2011aa} predicted that (i) the first-generation (FG) stars
in a GC can be at least 5--10 times more massive that the
second-generation (SG) stars at the epoch of GC formation and (ii) the
SG population is much more centrally concentrated than their FG
counterparts. This consequently leads to different two-body relaxation
timescales ($t_{\rm relax}$) for the two populations, because $t_{\rm
  relax}$ depends on the GC mass and size for a given typical stellar
mass. We emphasize that even though the observations indicate
  that approximately two-thirds of present-day cluster stars are SG
  descendants, with a substantial majority of FG stars having been
  stripped from their host clusters, it is still reasonable to discuss
  the effects of two-body relaxation on the dynamical evolution of a
  young 47 Tuc dominated by FG stars, because the stripping of FG
  stars takes hundreds of Myr. \citep{Bekki:2011aa}. We therefore
investigated the half-mass relaxation timescales\footnote{Here we use
  the median relaxation timescales \cite[e.g.,][]{Spitzer:1971aa} just
  for convenience in the discussion.} $t_{\rm relax}$ {\it separately}
for FG and SG stars to argue that dynamical mass segregation can
indeed happen for both stellar generations in 47 Tuc.

Figure \ref{Fig: timescale} shows $t_{\rm relax}$ as a function of the
initial half-mass radius\footnote{The observed half-mass radius of 47
  Tuc is $r_{\rm h} \approx 9.6 \pm 0.3$ pc (at the location of KA12's
  field), based on dynamical modeling \citep{Meylan:1989aa}, which is
  consistent with $r_{\rm h} = 10$ pc used in the recent simulations
  of \cite{Lane:2012aa}.} ($r_{\rm h}$) for FG and SG stars under the
assumptions that 47 Tuc is dominated by FG stars in the GC formation
scenario, where we assumed a typical stellar mass for 47 Tuc of
$m_\ast = 0.72 M_{\odot}$. We also assumed $M_{\rm SG}=5
  \times 10^5$ $M_{\odot}$ and $M_{\rm FG}=2.5 \times 10^6$
  $M_{\odot}$, based on the following arguments. The observed total
  mass of 47 Tuc is about $7 \times 10^5$ $M_{\odot}$
  \citep{forbes2010accreted}, and we assume that the observed fraction
  of SG stars in GCs is approximately 70\% (since we do not know
  exactly the current total mass of SG stars in 47 Tuc). The current
  total mass of SG stars is then approximately $5 \times 10^5$
  $M_{\odot}$. Previous theoretical models predict that the FG stars
  should be 5--10 times more massive than $M_{\rm SG}$, which implies
  that the initial total mass of the FG stars in the cluster ($M_{\rm
    FG}$) must be at least $2.5 \times 10^6$ $M_{\odot}$.
Consequently, we found that (i) $t_{\rm relax}$ of SG stars is less
than 13 Gyr for $r_{\rm h} \le 19.1$ pc. This means that mass
segregation of SG stars is possible for 47 Tuc within the observed
half-mass radius of 9.6 pc, (ii) $t_{\rm relax}$ of FG stars is less
than 13 Gyr for $r_{\rm h} \le 12.3$ pc. This implies that mass
segregation can also happen for the initial FG population of 47 Tuc,
provided that the initial FG population was as compact as the
cluster's present configuration. Even though the above argument is
oversimplified in the sense that we have estimated $t_{\rm relax}$
separately for FG and SG stars while in reality the two populations
will interact dynamically, this simple estimate has shown that
irrespective of the stellar generation, 47 Tuc must be dynamically
relaxed at its observed half-mass radius (9.6 pc), a region covered by
our {\sl HST} data.

As opposed to the inner region of 47 Tuc, the MF based on our VMC
data, which covers the radial range from $r = 500''$ to $r = 1100''$,
exhibits an almost constant power-law index:
see Fig. \ref{Fig: mass}. This observed slight decrease of the
$\alpha$ index out to radii of $r\in[500,600]''$, followed by the
absence of any radial trend, provides evidence againsts the importance
of mass segregation at these radii, which would have led to a gradual
increase of $\alpha$ with increasing radius (to at least 12.3 pc
$\equiv 550"$). Therefore, an additional dynamical effect must be
contributing to this unchanged $\alpha$ index. This might be due to
the effects of tidal stripping, in the sense that shock interactions
of 47 Tuc with the Galactic plane's gravitational potential can remove
low-mass stars from the cluster's periphery. \cite{Lane:2010aa} found
that 47 Tuc exhibits a rise in its local velocity dispersion at large
radii; subsequently \cite{Kupper:2010aa} found a similar increase in
the cluster's velocity dispersion in their $N$-body simulations, which
included the presence of the Galactic tidal
field. \cite{Kupper:2010aa} suggested that stars exhibiting abnormal
velocities at large radii may be ``potential escapers,'' which is an
indication of relaxation-driven evaporation, resulting in the transfer
of escaping stars from a cluster to its tidal tails
\citep{Fukushige:2000aa}. Based on this latter scenario, 47 Tuc is
expected to have cold tidal tails at relatively large radii in its
outer region \citep{Lane:2012aa}. Therefore, the overall picture that
is emerging on the basis of the observed MFs is that mass segregation
leads to high-mass stars moving to the cluster core, while ejecting
low-mass stars to its periphery. This is where tidal stripping occurs,
which removes the low-mass stars from the cluster through
gravitational interactions with the Milky Way's tidal field.

\begin{figure}[]
\begin{center}
\includegraphics[totalheight=0.25\textheight]{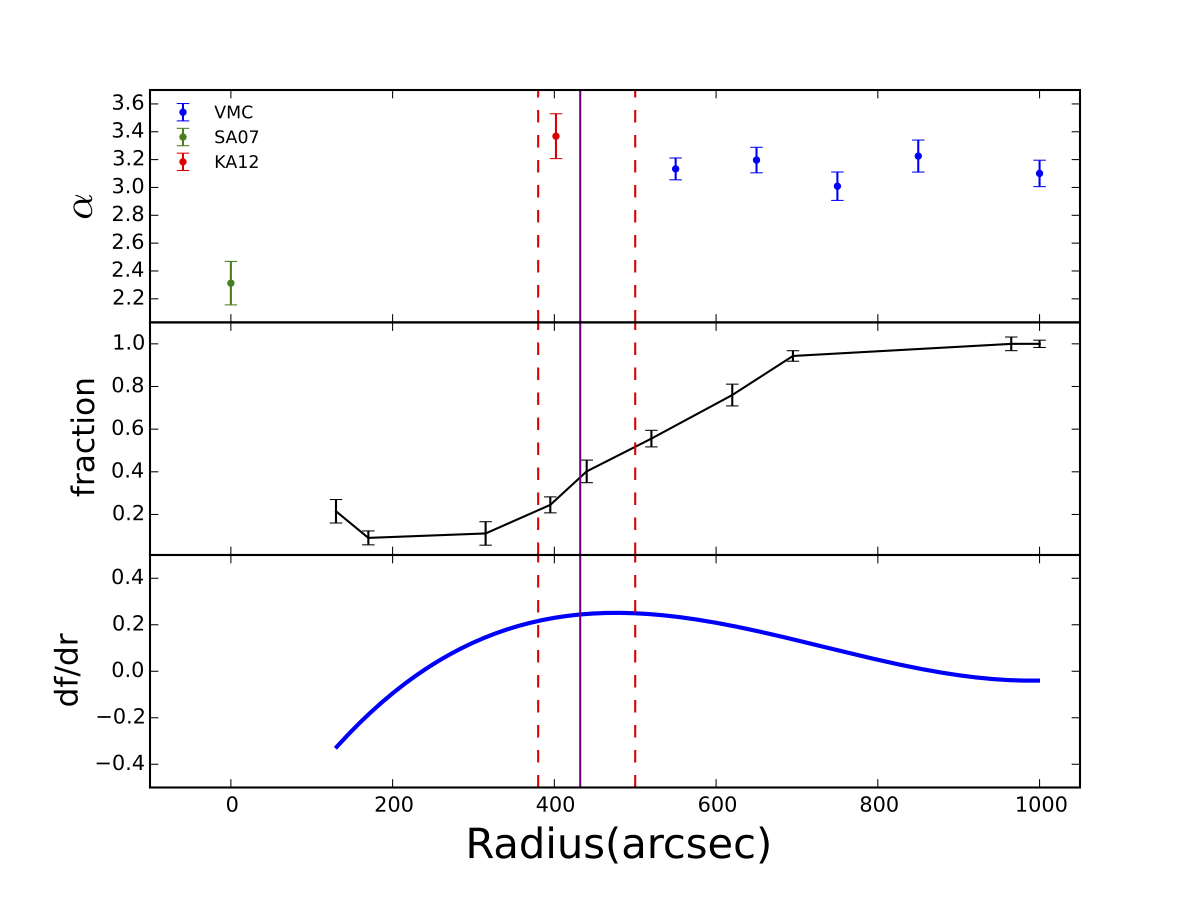}
\end{center}
\caption{Top: $\alpha$ as a function of radius in 47 Tuc.
 Middle: Fraction of FG stars, obtained by averaging the
  distributions of RGB and subgiant-branch stars in
  \cite{Li:2014aa}. Bottom: Derivative of the fraction of FG stars
  versus radius. The purple solid indicates the half-mass radius at
  about 9.6 pc ($430''$; we adopted a distance of 4.6 kpc to 47 Tuc)
  from the center of the cluster.}
\label{Fig: index}
\end{figure}

This overall picture can be further enriched by the presence of FG and
SG stars in 47 Tuc. \cite{Milone:2012ab} found that the fraction of
red red-giant-branch (RGB) stars increases from the cluster's
periphery (60\%) to its central regions (80\%). \cite{Li:2014aa} used
VMC data to show that the fraction of RGB and subgiant-branch stars in
the core is about 90\% and decreases to 10\% in the cluster's
periphery. We show all MF indices as a function of radius in
Fig. \ref{Fig: index} (top), together with the fraction of FG stars
(middle) and its derivative (bottom; smoothed). We note that $\alpha$
peaks for $r \in [380,500]''$, a region which is indicated by the two
red dashed lines. The $\alpha$ distribution then gradually becomes
constant. This behavior coincides remarkably well with that of the
derivative of the fraction of the FG of stars as a function of radius,
which peaks in the same radial range. The derivative of the
  fraction of FG stars with radius can be intuitively interpreted as
  the strength of the combined effects of mass segregation and tidal
  stripping as a function of radius, i.e., the closer to the core of
  the cluster one looks, the easier it is to observe the effects of
  mass segregation. At radii where mass segregation dominates, one
  would expect smaller fractions of FG stars to remain; similarly, the
  effects of tidal stripping start to dominate from radii closer to
  the cluster's half-mass radius outward. The fraction of FG stars is
  inversely correlated with the strength of the tidal stripping
  process.

But why is there also such a peak at radii $r \in [380,500]''$? We
suggest that this may be due to the competing effects of mass
segregation on the one hand and tidal stripping on the other.
The observed flattening of the MF toward the cluster core is
  caused by the FG massive stars having undergone mass segregation
  and, hence, they have migrated to the cluster center. This
  consequently increased the number of high-mass stars in the central
  regions (and, equivalently, increased the number of low-mass stars
  in the periphery). This process can proceed very rapidly in the
  cluster core, particularly if the core has initial substructure \citep{Kupper:2011aa}. SG stars will then form spatially more
  centrally from recycled FG stellar matter. Since the newly formed SG
  stars are more centrally concentrated, their spatial distribution
  would be relatively insensitive to the effects of tidal stripping.
On the other hand, in the outskirts of the cluster, both FG and SG
stars should be stripped tidally by the Milky Way's potential. The
stripping efficiency of the two generations is different, with the FG
being more efficiently stripped owing to its initially more diffuse
configuration. As a consequence, tidal stripping of low-mass FG stars
leads to a flatter $\alpha$ index toward larger radii, as indicated by
our VMC observations. The observed peak in the $\alpha$ distribution
is therefore likely caused by a combination of mass segregation and
tidal stripping operating at $r \in [380,500]''$.

\section{Conclusions}
\label{sec:conclusions}

In this paper, we have investigated the LFs and MFs of the low-mass MS
stars in 47 Tuc as a function of radius. We have used near-infrared
observations, obtained as part of the VMC survey, located at
clustercentric radii of 500--1100$''$, as well as data from two
complementary {\sl HST} data sets, one located in the central region
and the other at a distance of $6.7'$ from the cluster center. Our key
results are as follows.  \\

\begin{enumerate}
\item The {\sl HST}-based LFs and MFs suggest that the stellar numbers
  decline for magnitudes of $18.75\leq {\rm F606W}\leq 20.9$ in the
  center of the cluster, corresponding to a mass range of
  $0.55<m_\ast/{M_\odot}<0.73$. Contrary to this result, we found that
  the MF at $6.7'$ from the cluster center exhibits a deficit in stars
  at higher masses, $0.65<m_\ast/{M_\odot}< 0.82$. This is in support
  of the expectations from mass segregation, i.e., the high-mass stars
  sink to the center of the cluster and dynamically boost the
  lower-mass stars to the outer regions.
\item The 47 Tuc LFs and MFs based on the VMC VISTA data all exhibit
  power-law shapes in the mass range $0.55<m_\ast/{M_\odot}< 0.82$,
  with slopes remaining almost unchanged. This is likely due to tidal
  stripping, which is expected to remove the lower-mass FG stars from
  the outer regions of the cluster.
\item Combining the VMC and {\sl HST} observations, the MF index
  increases from the inner region outward and reaches a peak at a
  radius of about 380--500$''$; it subsequently decreases and remains
  constant toward the cluster's periphery. This peak in the $\alpha$
  distribution at radii $r \in [380,500]''$ (close to the half-mass
  radius) could be due to the combined effects of mass segregation and
  tidal stripping on the cluster's FG and SG stars.
\end{enumerate}

\begin{acknowledgements}
The analysis in this article is based on observations made with the
VISTA telescope at the European Southern Observatory under program ID
179.B-2003. We thank the team responsible for the UK's VISTA Data Flow
System for providing calibrated data products, supported by the UK's
Science and Technology Facilities Council. We also would like to thank
the Globular Cluster Treasury program team (PI: A. Sarajedini) for
their guidance on the {\sl HST} completeness table. C. Z. is grateful
to the Kavli Institute for Astronomy and Astrophysics for their
hospitality and dedicated assistance during the period of this work,
and special thanks are due to Yang Huang for his insightful
discussions on data reduction. C. Z., C. L., and R. d. G. acknowledge
funding support from the National Natural Science Foundation of China
(grant 11373010). C. Z., L. C. and C. L. acknowledge financial support from
``973 Program'' 2014 CB845702 and the Strategic Priority Research
Program ``The Emergence of Cosmological Structures'' of the Chinese
Academy of Sciences (CAS; grants XDB09010100 and XDB09000000).

\end{acknowledgements}


\begin{thebibliography}{42}
\expandafter\ifx\csname natexlab\endcsname\relax\def\natexlab#1{#1}\fi

\bibitem[{{Anderson} \& {King}(1996)}]{Anderson:1996aa}
{Anderson}, J., \& {King}, I.~R. 1996, in Astronomical Society of the Pacific
  Conference Series, Vol.~92, Formation of the Galactic Halo...Inside and Out,
  ed. H.~L. {Morrison} \& A.~{Sarajedini}, 257

\bibitem[{{Anderson} {et~al.}(2009){Anderson}, {Piotto}, {King}, {Bedin}, \&
  {Guhathakurta}}]{Anderson:2009aa}
{Anderson}, J., {Piotto}, G., {King}, I.~R., {Bedin}, L.~R., \& {Guhathakurta},
  P. 2009, \apjl, 697, L58

\bibitem[{{Anderson} {et~al.}(2008){Anderson}, {Sarajedini}, {Bedin}, {King},
  {Piotto}, {Reid}, {Siegel}, {Majewski}, {Paust}, {Aparicio}, {Milone},
  {Chaboyer}, \& {Rosenberg}}]{Anderson:2008aa}
{Anderson}, J., {Sarajedini}, A., {Bedin}, L.~R., {et~al.} 2008, \aj, 135, 2055

\bibitem[{{Bekki}(2011)}]{Bekki:2011aa}
{Bekki}, K. 2011, \mnras, 412, 2241

\bibitem[{{Bochanski} {et~al.}(2010){Bochanski}, {Hawley}, {Covey}, {West},
  {Reid}, {Golimowski}, \& {Ivezi{\'c}}}]{Bochanski:2010aa}
{Bochanski}, J.~J., {Hawley}, S.~L., {Covey}, K.~R., {et~al.} 2010, \aj, 139,
  2679

\bibitem[{{Bonnell} {et~al.}(2007){Bonnell}, {Larson}, \&
  {Zinnecker}}]{Bonnell:2007aa}
{Bonnell}, I.~A., {Larson}, R.~B., \& {Zinnecker}, H. 2007, Protostars and
  Planets V, 149

\bibitem[{{Cioni} {et~al.}(2011){Cioni}, {Clementini}, {Girardi}, {Guandalini},
  {Gullieuszik}, {Miszalski}, {Moretti}, {Ripepi}, {Rubele}, {Bagheri},
  {Bekki}, {Cross}, {de Blok}, {de Grijs}, {Emerson}, {Evans}, {Gibson},
  {Gonzales-Solares}, {Groenewegen}, {Irwin}, {Ivanov}, {Lewis}, {Marconi},
  {Marquette}, {Mastropietro}, {Moore}, {Napiwotzki}, {Naylor}, {Oliveira},
  {Read}, {Sutorius}, {van Loon}, {Wilkinson}, \& {Wood}}]{Cioni:2011aa}
{Cioni}, M.-R.~L., {Clementini}, G., {Girardi}, L., {et~al.} 2011, \aap, 527,
  A116

\bibitem[{{de Grijs} {et~al.}(2002){de Grijs}, {Gilmore}, {Johnson}, \&
  {Mackey}}]{de-Grijs:2002aa}
{de Grijs}, R., {Gilmore}, G.~F., {Johnson}, R.~A., \& {Mackey}, A.~D. 2002,
  \mnras, 331, 245

\bibitem[{{de Marchi} \& {Paresce}(1995)}]{de-Marchi:1995aa}
{de Marchi}, G., \& {Paresce}, F. 1995, \aap, 304, 211

\bibitem[{{D'Ercole} {et~al.}(2010){D'Ercole}, {D'Antona}, {Ventura},
  {Vesperini}, \& {McMillan}}]{DErcole:2010aa}
{D'Ercole}, A., {D'Antona}, F., {Ventura}, P., {Vesperini}, E., \& {McMillan},
  S.~L.~W. 2010, \mnras, 407, 854

\bibitem[{{Dotter} {et~al.}(2008){Dotter}, {Chaboyer}, {Jevremovi{\'c}},
  {Kostov}, {Baron}, \& {Ferguson}}]{Dotter:2008aa}
{Dotter}, A., {Chaboyer}, B., {Jevremovi{\'c}}, D., {et~al.} 2008, \apjs, 178,
  89

\bibitem[{Forbes \& Bridges(2010)}]{forbes2010accreted}
Forbes, D.~A., \& Bridges, T. 2010, Monthly Notices of the Royal Astronomical
  Society, 404, 1203

\bibitem[{{Fukushige} \& {Heggie}(2000)}]{Fukushige:2000aa}
{Fukushige}, T., \& {Heggie}, D.~C. 2000, \mnras, 318, 753

\bibitem[{{Giersz} \& {Heggie}(2011)}]{Giersz:2011aa}
{Giersz}, M., \& {Heggie}, D.~C. 2011, \mnras, 410, 2698

\bibitem[{{Girardi} {et~al.}(2010){Girardi}, {Williams}, {Gilbert},
  {Rosenfield}, {Dalcanton}, {Marigo}, {Boyer}, {Dolphin}, {Weisz},
  {Melbourne}, {Olsen}, {Seth}, \& {Skillman}}]{Girardi:2010aa}
{Girardi}, L., {Williams}, B.~F., {Gilbert}, K.~M., {et~al.} 2010, \apj, 724,
  1030

\bibitem[{{Harris}(1996)}]{Harris:1996aa}
{Harris}, W.~E. 1996, \aj, 112, 1487

\bibitem[{{Hesser} {et~al.}(1987){Hesser}, {Harris}, {Vandenberg}, {Allwright},
  {Shott}, \& {Stetson}}]{Hesser:1987aa}
{Hesser}, J.~E., {Harris}, W.~E., {Vandenberg}, D.~A., {et~al.} 1987, \pasp,
  99, 739

\bibitem[{{Heyl} {et~al.}(2015){Heyl}, {Richer}, {Antolini}, {Goldsbury},
  {Kalirai}, {Parada}, \& {Tremblay}}]{Heyl:2015aa}
{Heyl}, J., {Richer}, H.~B., {Antolini}, E., {et~al.} 2015, \apj, 804, 53

\bibitem[{{Irwin} {et~al.}(2004){Irwin}, {Lewis}, {Hodgkin}, {Bunclark},
  {Evans}, {McMahon}, {Emerson}, {Stewart}, \& {Beard}}]{Irwin:2004aa}
{Irwin}, M.~J., {Lewis}, J., {Hodgkin}, S., {et~al.} 2004, in Society of
  Photo-Optical Instrumentation Engineers (SPIE) Conference Series, Vol. 5493,
  Optimizing Scientific Return for Astronomy through Information Technologies,
  ed. P.~J. {Quinn} \& A.~{Bridger}, 411--422

\bibitem[{{Kalirai} {et~al.}(2012){Kalirai}, {Richer}, {Anderson}, {Dotter},
  {Fahlman}, {Hansen}, {Hurley}, {King}, {Reitzel}, {Rich}, {Shara}, {Stetson},
  \& {Woodley}}]{Kalirai:2012aa}
{Kalirai}, J.~S., {Richer}, H.~B., {Anderson}, J., {et~al.} 2012, \aj, 143, 11

\bibitem[{{Kalirai} {et~al.}(2013){Kalirai}, {Anderson}, {Dotter}, {Richer},
  {Fahlman}, {Hansen}, {Hurley}, {Reid}, {Rich}, \& {Shara}}]{Kalirai:2013aa}
{Kalirai}, J.~S., {Anderson}, J., {Dotter}, A., {et~al.} 2013, \apj, 763, 110

\bibitem[{{King}(1958)}]{King:1958aa}
{King}, I. 1958, \aj, 63, 465

\bibitem[{{Kroupa} {et~al.}(1990){Kroupa}, {Tout}, \&
  {Gilmore}}]{Kroupa:1990aa}
{Kroupa}, P., {Tout}, C.~A., \& {Gilmore}, G. 1990, \mnras, 244, 76

\bibitem[{{K{\"u}pper} {et~al.}(2010){K{\"u}pper}, {Kroupa}, {Baumgardt}, \&
  {Heggie}}]{Kupper:2010aa}
{K{\"u}pper}, A.~H.~W., {Kroupa}, P., {Baumgardt}, H., \& {Heggie}, D.~C. 2010,
  \mnras, 407, 2241

\bibitem[{{K{\"u}pper} {et~al.}(2011){K{\"u}pper}, {Maschberger}, {Kroupa}, \&
  {Baumgardt}}]{Kupper:2011aa}
{K{\"u}pper}, A.~H.~W., {Maschberger}, T., {Kroupa}, P., \& {Baumgardt}, H.
  2011, \mnras, 417, 2300

\bibitem[{{Lane} {et~al.}(2012){Lane}, {K{\"u}pper}, \& {Heggie}}]{Lane:2012aa}
{Lane}, R.~R., {K{\"u}pper}, A.~H.~W., \& {Heggie}, D.~C. 2012, \mnras, 423,
  2845

\bibitem[{{Lane} {et~al.}(2010){Lane}, {Brewer}, {Kiss}, {Lewis}, {Ibata},
  {Siebert}, {Bedding}, {Sz{\'e}kely}, \& {Szab{\'o}}}]{Lane:2010aa}
{Lane}, R.~R., {Brewer}, B.~J., {Kiss}, L.~L., {et~al.} 2010, \apjl, 711, L122

\bibitem[{{Larson}(1992)}]{Larson:1992aa}
{Larson}, R.~B. 1992, \mnras, 256, 641

\bibitem[{{Li} {et~al.}(2014){Li}, {de Grijs}, {Deng}, {Rubele}, {Wang},
  {Bekki}, {Cioni}, {Clementini}, {Emerson}, {For}, {Girardi}, {Groenewegen},
  {Guandalini}, {Gullieuszik}, {Marconi}, {Piatti}, {Ripepi}, \& {van
  Loon}}]{Li:2014aa}
{Li}, C., {de Grijs}, R., {Deng}, L., {et~al.} 2014, \apj, 790, 35

\bibitem[{{Marigo} {et~al.}(2008){Marigo}, {Girardi}, {Bressan}, {Groenewegen},
  {Silva}, \& {Granato}}]{Marigo:2008aa}
{Marigo}, P., {Girardi}, L., {Bressan}, A., {et~al.} 2008, \aap, 482, 883

\bibitem[{{Meylan}(1989)}]{Meylan:1989aa}
{Meylan}, G. 1989, \aap, 214, 106

\bibitem[{{Milone} {et~al.}(2012){Milone}, {Piotto}, {Bedin}, {King},
  {Anderson}, {Marino}, {Bellini}, {Gratton}, {Renzini}, {Stetson}, {Cassisi},
  {Aparicio}, {Bragaglia}, {Carretta}, {D'Antona}, {Di Criscienzo},
  {Lucatello}, {Monelli}, \& {Pietrinferni}}]{Milone:2012ab}
{Milone}, A.~P., {Piotto}, G., {Bedin}, L.~R., {et~al.} 2012, \apj, 744, 58

\bibitem[{{Padoan} \& {Nordlund}(2002)}]{Padoan:2002aa}
{Padoan}, P., \& {Nordlund}, {\AA}. 2002, \apj, 576, 870

\bibitem[{{Paust} {et~al.}(2010){Paust}, {Reid}, {Piotto}, {Aparicio},
  {Anderson}, {Sarajedini}, {Bedin}, {Chaboyer}, {Dotter}, {Hempel},
  {Majewski}, {Mar{\'{\i}}n-Franch}, {Milone}, {Rosenberg}, \&
  {Siegel}}]{Paust:2010aa}
{Paust}, N.~E.~Q., {Reid}, I.~N., {Piotto}, G., {et~al.} 2010, \aj, 139, 476

\bibitem[{{Pietrinferni} {et~al.}(2006){Pietrinferni}, {Cassisi}, {Salaris}, \&
  {Castelli}}]{Pietrinferni:2006aa}
{Pietrinferni}, A., {Cassisi}, S., {Salaris}, M., \& {Castelli}, F. 2006, \apj,
  642, 797

\bibitem[{{Rubele} {et~al.}(2012){Rubele}, {Kerber}, {Girardi}, {Cioni},
  {Marigo}, {Zaggia}, {Bekki}, {de Grijs}, {Emerson}, {Groenewegen},
  {Gullieuszik}, {Ivanov}, {Miszalski}, {Oliveira}, {Tatton}, \& {van
  Loon}}]{Rubele:2012aa}
{Rubele}, S., {Kerber}, L., {Girardi}, L., {et~al.} 2012, \aap, 537, A106

\bibitem[{{Rubele} {et~al.}(2015){Rubele}, {Girardi}, {Kerber}, {Cioni},
  {Piatti}, {Zaggia}, {Bekki}, {Bressan}, {Clementini}, {de Grijs}, {Emerson},
  {Groenewegen}, {Ivanov}, {Marconi}, {Marigo}, {Moretti}, {Ripepi},
  {Subramanian}, {Tatton}, \& {van Loon}}]{Rubele:2015aa}
{Rubele}, S., {Girardi}, L., {Kerber}, L., {et~al.} 2015, \mnras, 449, 639

\bibitem[{{Santiago} {et~al.}(1996){Santiago}, {Elson}, \&
  {Gilmore}}]{Santiago:1996aa}
{Santiago}, B.~X., {Elson}, R.~A.~W., \& {Gilmore}, G.~F. 1996, \mnras, 281,
  1363

\bibitem[{{Sarajedini} {et~al.}(2007){Sarajedini}, {Bedin}, {Chaboyer},
  {Dotter}, {Siegel}, {Anderson}, {Aparicio}, {King}, {Majewski},
  {Mar{\'{\i}}n-Franch}, {Piotto}, {Reid}, \& {Rosenberg}}]{Sarajedini:2007aa}
{Sarajedini}, A., {Bedin}, L.~R., {Chaboyer}, B., {et~al.} 2007, \aj, 133, 1658

\bibitem[{{Spitzer} \& {Hart}(1971)}]{Spitzer:1971aa}
{Spitzer}, Jr., L., \& {Hart}, M.~H. 1971, \apj, 166, 483

\bibitem[{{Stetson}(1987)}]{Stetson:1987aa}
{Stetson}, P.~B. 1987, \pasp, 99, 191

\bibitem[{{Valcarce} {et~al.}(2012){Valcarce}, {Catelan}, \&
  {Sweigart}}]{Valcarce:2012aa}
{Valcarce}, A.~A.~R., {Catelan}, M., \& {Sweigart}, A.~V. 2012, \aap, 547, A5

\end{thebibliography}

\end{document}